\documentclass[aps,prl,twocolumn,groupedaddress,showpacs,showkeys]{revtex4-1}
\usepackage{amsfonts}
\usepackage{mathtools}
\usepackage{amsmath}
\usepackage{amssymb}
\usepackage{float}
\usepackage{subdepth}
\usepackage{color}
\usepackage{graphicx}
\usepackage{natbib}
\usepackage{hyperref}
\usepackage{url}
\usepackage{soul}
\hypersetup{colorlinks, linkcolor={blue}, citecolor={blue}, urlcolor={blue}}

\begin{document}
\title{The Ferroelectric Point Contact}
\author{Ping Tang$^{1}$}
\author{Ryo Iguchi$^{2}$}
\author{Ken-ichi Uchida$^{2,3,4}$}
\author{Gerrit E. W. Bauer$^{1,3,4,5}$}
\affiliation{$^1$WPI-AIMR, Tohoku
University, 2-1-1 Katahira, 980-8577 Sendai, Japan}
\affiliation{$^2$National Institute for Materials Science, Tsukuba 305-0047, Japan}
\affiliation{$^3$Institute for Materials Research, Tohoku University, 2-1-1 Katahira, 980-8577 Sendai, Japan}
\affiliation{$^4$Center for Spintronics Research Network, Tohoku University, Sendai 980-8577, Japan}
\affiliation{$^5$Zernike Institute for Advanced Materials, University of Groningen, 9747 AG Groningen, Netherlands}

\begin{abstract}
We formulate a scattering theory of polarization and heat transport through a
ballistic ferroelectric point contact. We predict a polarization current under
either an electric field or a temperature difference that depends strongly on
the direction of the ferroelectric order and can be detected by its magnetic
stray field and associated thermovoltage and Peltier effect.
\end{abstract}
\maketitle




Orifices such as Maxwell/Sharvin point contacts
\cite{maxwell1873treatise,sharvin1965possible}, micro- and nano-structured
constrictions such as semiconductor quantum point contacts
\cite{van1992thermo,van1996quantum} and atomic-scale break junctions
\cite{agrait2003quantum} etc. are important instruments to study transport
properties in condensed matter physics on small length scales. The
quantization of transport of electrons \cite{van1988quantized,wharam1988one},
light \cite{montie1991observation}, super-
\cite{beenakker1991josephson,takayanagi1995observation}, and spin
\cite{debray2009all,meier2003magnetization} currents as well as photonic
\cite{meschke2006single,ojanen2007photon}\textit{,} electronic
\cite{chiatti2006quantum,jezouin2013quantum,cui2017quantized} and phononic
\cite{rego1998quantized,blencowe1999quantum,schwab2000measurement} heat
currents are some of the important breakthroughs in this field. To the best of
our knowledge, the transport through constrictions formed by
\emph{ferroelectrics} has never been addressed, neither theoretically nor experimentally.

Ferroelectricity refers to the electrically switchable macroscopic order of
electric dipoles or persistent polarization that forms spontaneously below a
Curie temperature \cite{xu2013ferroelectric} with some analogies with
magnetism \cite{spaldin2007analogies}. In so-called displacive ferroelectrics,
the phase transition is accompanied by a symmetry-breaking structural phase
transition: ferroelectricity is then caused by short-range elastic forces,
while magnetism is caused by the short-range exchange interaction. The dipolar
interaction is much larger in ferroelectrics than in ferromagnets and causes
important secondary effects.

A magnon, the quantum of a spin wave, carries energy, linear, and spin angular
momentum. In high quality magnetic insulators, coherently excited magnons with
long wave lengths travel ballistically over large distances
\cite{kajiwara2010transmission}. Thermal magnons, on the other hand, propagate
diffusely under a gradient of temperature, magnetic field, or chemical
potential \cite{cornelissen2016magnon} and cause the spin Seebeck
\cite{uchida2010spin} and spin Peltier effects
\cite{flipse2014observation,daimon2016thermal}. We previously addressed the
diffuse polarization transport in ferroelectric capacitors
\cite{PhysRevLett.126.187603}.

In this Letter, we formulate ballistic transport through constrictions of a
displacive ferroelectric by the scattering theory of transport
\cite{buttiker1992scattering,datta1997electronic,blencowe1999quantum}. In
contrast to the findings for diffuse systems \cite{PhysRevLett.126.187603}, we
predict a dc polarization current that generates observable stray magnetic
fields as well as a dc polarization Peltier effect.

\begin{figure}[ptb]
\centering
\par
\includegraphics[width=7.2 cm]{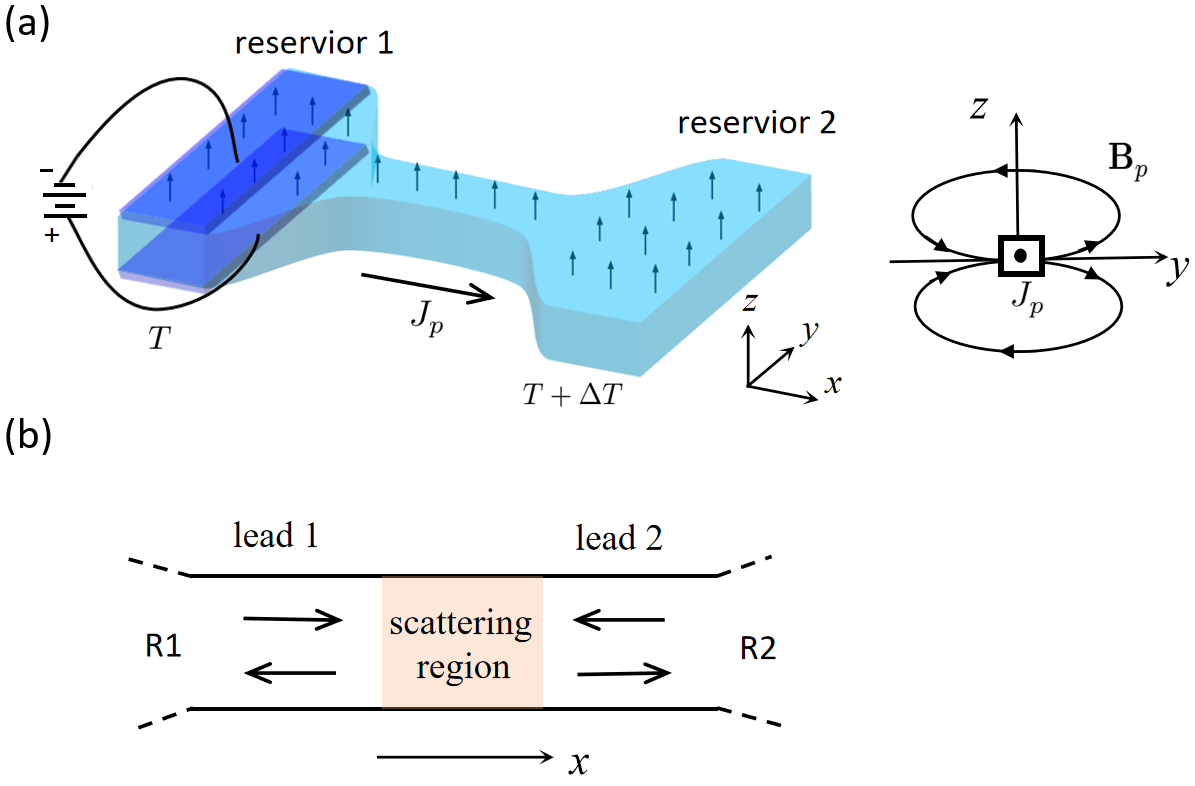}\newline\caption{(a) Polarization
transport between two ferroelectric reservoirs (R1 and R2) connected by a
ferroelectric lead (left panel). The polarization current induces a magnetic
field (right panel). In accordance with the polarization current direction as
in (a), there are an electric field applied on the R1 along the direction of
ferroelectric order and a temperature difference ($\Delta T>0$) on the R2. (b)
The ferroelectric lead is divided by the scattering region into two parts of
lead 1 and 2 that are in adiabatic contact with the R1 and R2, respectively.}%
\label{Fig-1}%
\end{figure}

We consider a monolithic ferroelectric sample in which a narrow wire
adiabatically connects two reservoirs with perpendicular ferroelectric order
at temperatures sufficiently below the phase transition. The two reservoirs
are at thermal equilibrium but at possibly different temperature $T_{1}\ $and
$T_{2}.$ Top and bottom gates allow application of different electric fields
$E_{1}\ $and $E_{2}$ as sketched in Fig. ~\ref{Fig-1}(a). We focus on
steady-state transport, which requires sufficiently large reservoirs. In
linear response, currents and forces are related by a matrix of transport
coefficients \cite{PhysRevLett.126.187603}
\begin{equation}
\left(
\begin{matrix}
-J_{p}\\
J_{q}%
\end{matrix}
\right)  =G\left(
\begin{matrix}
1 & S\\
\Pi & K/G
\end{matrix}
\right)  \left(
\begin{matrix}
\Delta E\\
-\Delta T
\end{matrix}
\right)  \label{resp}%
\end{equation}
where the $J_{p}$ and $J_{q}$ are the polarization and heat currents flowing
from reservoir 1 to 2, while $\Delta E=E_{2}-E_{1}$ and $\Delta T=T_{2}-T_{1}$
are the electric field and temperature differences between the two reservoirs,
respectively. $G$ is the polarization conductance, $S$ ($\Pi$) the
polarization Seebeck (Peltier) coefficient with Kelvin-Onsager relation
$\Pi=ST$, and $K$ the thermal conductance. We now derive all transport
coefficients by the scattering theory of transport.

\emph{Phonon model:} A phonon mode in a wire along the $x$-axis with wave
number $k$ reads
\begin{equation}
{\boldsymbol{u}}_{nk\sigma}(l,s)=\frac{1}{\sqrt{NM_{s}}}\mathbf{e}_{nk\sigma
}(s)\varphi_{n}({\boldsymbol{\rho}}_{l})e^{ikx_{l}-i\omega_{nk\sigma}\tau}
\label{pnm}%
\end{equation}
where ${\boldsymbol{u}}_{nk\sigma}(l,s)$ is the displacement of the $s$-th ion
in the $l$-th ferroelectric unit cell, with the ionic mass of $M_{s}$, $N$ the
number of unit cells. Here $\omega_{nk\sigma}$ is the phonon frequency
dispersion, $\mathbf{e}_{nk\sigma}(s)$ the polarization vector of phonon with
polarization index $\sigma=1,\cdots,3m,$ where $m$ is the number of ions in
each ferroelectric unit cell, with band index $n$ and ${\boldsymbol{\rho}}%
_{l}$ the transverse coordinate. The orthogonality relations are $\sum
_{l}\varphi_{n}({\boldsymbol{\rho}}_{l})\varphi_{n^{\prime}}({\boldsymbol{\rho
}}_{l})^{\ast}e^{i(k-k^{\prime})x_{l}}=N\delta_{nn^{\prime}}\delta
_{kk^{\prime}}$ and $\sum_{s}\mathbf{e}_{nk\sigma}(s)\cdot\lbrack
\mathbf{e}_{nk\sigma^{\prime}}(s)]^{\ast}=\delta_{\sigma\sigma^{\prime}}$.

A phonon ${\boldsymbol{u}}_{nk\sigma}$ originating in the left reservoir (R1)
with positive group velocity $v_{nk\sigma}=\partial\omega_{nk\sigma}/\partial
k>0$ can be elastically reflected or transmitted by the constriction, as
illustrated in Fig.~\ref{Fig-1}(b), and the amplitude in both leads then
reads
\begin{equation}
{\boldsymbol{u}}_{nk\sigma}^{(1)}=\left\{
\begin{array}
[c]{c}%
{\boldsymbol{u}}_{nk\sigma}+\sum_{n^{\prime},\sigma^{\prime}}{\boldsymbol{u}%
}_{n^{\prime}-k^{\prime}\sigma^{\prime}}r_{n^{\prime}\sigma^{\prime}n\sigma}\\
\sum_{n^{\prime},\sigma^{\prime}}{\boldsymbol{u}}_{n^{\prime}k^{\prime}%
\sigma^{\prime}}t_{n^{\prime}\sigma^{\prime}n\sigma}%
\end{array}
\;\mathrm{in}\;%
\begin{array}
[c]{c}%
\mathrm{lead\;1}\\
\mathrm{lead\;2}%
\end{array}
\right.  ,
\end{equation}
where $r_{n^{\prime}\sigma^{\prime}n\sigma}$ is the reflection amplitude from
mode $n\sigma$ to $n^{\prime}\sigma^{\prime}$ in lead 1, while $t_{n^{\prime
}\sigma^{\prime}n\sigma}$ is the transmission amplitude from mode $n\sigma$ in
lead 1 to $n^{\prime}\sigma^{\prime}$ in lead 2 and $\omega_{nk\sigma}%
=\omega_{n^{\prime}k^{\prime}\sigma^{\prime}}$ for the elastic scattering.
Analogously, assuming that the junction is symmetric, an incoming phonon
${\boldsymbol{u}}_{nk\sigma}$ from the right reservoir (R2) generates the
following amplitudes in both leads
\begin{equation}
{\boldsymbol{u}}_{n-k\sigma}^{(2)}=\left\{
\begin{array}
[c]{c}%
\sum_{n^{\prime},\sigma^{\prime}}{\boldsymbol{u}}_{n^{\prime}-k^{\prime}%
\sigma^{\prime}}t_{n^{\prime}\sigma^{\prime}n\sigma}\\
{\boldsymbol{u}}_{n-k\sigma}+\sum_{n^{\prime},\sigma^{\prime}}{\boldsymbol{u}%
}_{n^{\prime}k^{\prime}\sigma^{\prime}}r_{n^{\prime}\sigma^{\prime}n\sigma}%
\end{array}
\;\mathrm{in}\;%
\begin{array}
[c]{c}%
\mathrm{lead\;1}\\
\mathrm{lead\;2}%
\end{array}
\right.  .
\end{equation}
The total displacement field operator in the leads can then be written as
\cite{blencowe1999quantum}
\begin{align}
\hat{\boldsymbol{u}}(l,s)  &  =\sum_{n,\sigma,k>0}\sqrt{\frac{\hbar}%
{2\omega_{nk\sigma}}}\left[  \hat{a}_{nk\sigma}^{(1)}{\boldsymbol{u}%
}_{nk\sigma}^{(1)}(l,s)+\right. \nonumber\\
&  \left.  \hat{a}_{n-k\sigma}^{(2)}{\boldsymbol{u}}_{n-k\sigma}%
^{(2)}(l,s)\right]  +\mathrm{h.c.} \label{disoper}%
\end{align}
where $\hat{a}_{nk\sigma}^{(1)}$ ($\hat{a}_{n-k\sigma}^{(2)}$) represents the
creation operator of phonons originating in R1 (R2) and h.c. denotes the
hermitian conjugate of the previous term.

We adopt the standard rigid-ion approximation for displacive ferroelectrics
\cite{born1954dynamical,freire1988lattice} with polarization fluctuations
around the ground state,
\begin{equation}
\delta\hat{\mathbf{P}}_{l}=\sum_{s}Q_{s}\hat{\boldsymbol{u}}(l,s),\label{PFL}%
\end{equation}
where $Q_{s}$ is the ionic charge of the $s$-th ion in the ferroelectric unit
cell with index $l$ and $\sum_{s}Q_{s}=0$. The polarization projection
along the ferroelectric order in the $l$-th unit cell $\hat{P}_{l}=\left(
P_{0}+\triangle\hat{P}_{l}^{\parallel}\right)  \cos\triangle\hat{\theta}_{l}$,
where $\triangle\hat{P}_{l}^{\parallel}$ is the oscillation of the
polarization modulus relative to $P_{0}=\left\vert \mathbf{P}%
 _{0}\right\vert $ in the ground state and $\triangle\hat{\theta
}_{l}$ a small angle of rotation. We now make the \textit{ferron} approximation that the fluctuations in $\triangle\hat{P}_{l}^{\parallel}$ and $\triangle
\hat{\theta}_{l}$ are uncorrelated, which is justified when the longitudinal
elastic constant of the dipole is sufficiently larger than the transverse one
and excellent for the order-disorder type of ferroelectrics with stable
molecular dipoles \citep{pouget1986lattice}. To leading order in small
\textit{Cartesian} transverse and longitudinal fluctuations, i.e., $\triangle
\hat{\mathbf{P}}_{l}^{\perp}=\triangle\hat{\mathbf{P}}_{l}-\left(
\triangle\hat{\mathbf{P}}_{l}\cdot\mathbf{P}_{0}\right)/P_{0}%
$ and $\delta\hat{P}_{l}^{\parallel}=\triangle\hat{\mathbf{P}}_{l}\cdot
\mathbf{P}_{0}/P_{0}$,
\begin{equation}
\hat{P}_{l}=P_{0}\left[  1-\frac{(\triangle\hat{\mathbf{P}}_{l}^{\perp})^{2}%
}{2P_{0}^{2}}\right]  +\triangle\hat{P}_{l}^{\parallel}+\mathcal{O}%
[(\triangle\hat{\mathbf{P}}_{l}^{\perp})^{2},\triangle\hat{P}_{l}^{\parallel
}].\label{PFE}%
\end{equation}
The first term captures the reduced polarization projection along the
ferroelectric order by the transverse fluctuations and we disregard
$\triangle\hat{P}_{l}^{\parallel}$ and higher-order terms.

When the polarization is conserved on the length scale of the constriction,
the operator for the coarse-grained polarization density $\hat{P}_{l}%
/\Omega\longrightarrow{\hat{p}}\left(  \mathbf{r},\tau\right)  $ per unit cell
with volume $\Omega$ is related to the polarization current density
$\hat{\jmath}_{p}(\mathbf{r},\tau)$ through
\begin{equation}
\partial_{x}\hat{\mathbf{\jmath}}_{p}(\mathbf{r},\tau)=-\partial_{\tau}%
{\hat{p}}(\mathbf{r},\tau),\label{conservation}%
\end{equation}
where $\mathbf{r}\equiv(x,{\boldsymbol{\rho}})$ and the operators are in the
Heisenberg picture. The partial Fourier transform $\hat{p}({\boldsymbol{\rho}%
},q,\omega)=\int d\tau\int dx{\hat{p}}({\boldsymbol{\rho}},x,\tau
)e^{i\omega\tau-iqx}$ leads to
\begin{equation}
\hat{\mathbf{\jmath}}_{p}(\mathbf{r},\tau)=\int\frac{dq}{2\pi}\int%
\frac{d\omega}{2\pi}\left(  \frac{\omega}{q}\right)  \hat{p}(\boldsymbol{\rho
},q,\omega)e^{-i\omega\tau+iqx}.\label{partialfourier}%
\end{equation}
Substituting Eq.~(\ref{PFE}) and to leading order in the fluctuations
Eq.~(\ref{PFL}), the statistical average of the total polarization current
\begin{align}
J_{p} &  \equiv\int d{\boldsymbol{\rho}}\,\langle\mathbf{\hat{\jmath}}%
_{p}(\mathbf{r},\tau)\rangle\nonumber\\
&  =-\frac{\hbar}{2P_{0}}\sum_{n^{\prime},\sigma^{\prime}}\sum_{n,\sigma}%
\int_{0}^{\mathrm{BZ}}\frac{dk}{2\pi}\frac{|{\boldsymbol{F}}_{n^{\prime
}k^{\prime}\sigma^{\prime}}^{\perp}|^{2}}{\omega_{n^{\prime}k^{\prime}%
\sigma^{\prime}}}v_{n^{\prime}k^{\prime}\sigma^{\prime}}|t_{n^{\prime}%
\sigma^{\prime}n\sigma}|^{2}\nonumber\\
&  \times\left[  \langle\hat{a}_{nk\sigma}^{\dagger}\hat{a}_{nk\sigma}%
\rangle^{{(1)}}-\langle\hat{a}_{n-k\sigma}^{\dagger}\hat{a}_{n-k\sigma}%
\rangle^{{(2)}}\right]  ,\label{current}%
\end{align}
where $\mathrm{BZ}$ indicates the Brillouin zone boundary. Here
${\boldsymbol{F}}_{nk\sigma}^{\perp}={\boldsymbol{F}}_{nk\sigma}%
-\left({\boldsymbol{F}}_{nk\sigma}\cdot \mathbf{P}_{0}\right)/P_{0}$ is the transverse component of
\begin{equation}
{\boldsymbol{F}}_{nk\sigma}=\sum_{s}\frac{Q_{s}}{\sqrt{M_{s}}}{\mathbf{e}%
}_{nk\sigma}(s)\
\end{equation}
and $\langle\cdots\rangle^{{(i)}}$ is a thermal average in reservoir $i:$
\begin{equation}
\langle\hat{a}_{nk\sigma}^{\dagger}\hat{a}_{nk\sigma}\rangle^{{(i)}}=N\left(
\hbar\omega_{nk\sigma}-\mu_{nk\sigma}^{\left(  i\right)  },T^{\left(
i\right)  }\right)  .
\end{equation}
The parameters in the Planck distribution function $N(\hbar\omega
,T)\equiv1/[e^{\hbar\omega/k_{B}T}-1]$ are the temperature $T^{\left(
i\right)  }$ and effective chemical potential $\mu_{nk\sigma}^{\left(
i\right)  }=\xi_{nk\sigma}E^{\left(  i\right)  }$, where $E^{\left(  i\right)
}$ is the electric field of reservoir $i$ and polarization
\begin{equation}
\xi_{nk\sigma}=-\frac{\hbar}{2P_{0}}\frac{|{\boldsymbol{F}}_{nk\sigma}^{\perp
}|^{2}}{\omega_{nk\sigma}}.\label{Polph}%
\end{equation}

In the above calculations we adopted the conventional assumption of an
adiabatic connection between the reservoirs and the leads. The electric fields
are applied on the reservoirs where they affect the equilibrium phonon
statistics. In linear response, transport is then only governed by the
differences in $E^{\left(  i\right)  }$ and $T^{\left(  i\right)  }$ in the
reservoirs and the transmission coefficients through the constriction. We
consider here two different configurations, viz. with equilibrium polarization
in the constriction normal to the plane (parallel to those in the reservoirs)
and along the wire. They can in principle be switched by the electric fields
of local gates, but we assume here for simplicity that the field in the
constriction vanishes. 

The energy or heat current reads
\begin{align}
J_{q} &  =\sum_{n^{\prime}\sigma^{\prime}}\sum_{n\sigma}\int_{0}^{\mathrm{BZ}%
}\frac{dk}{2\pi}\hbar\omega_{n^{\prime}k^{\prime}\sigma^{\prime}}v_{n^{\prime
}k^{\prime}\sigma^{\prime}}|t_{n^{\prime}\sigma^{\prime}n\sigma}%
|^{2}\nonumber\\
&  \times\left[  N\left(  \hbar\omega_{nk\sigma}-\mu_{nk\sigma}^{\left(
1\right)  },T^{\left(  1\right)  }\right)  \right.  \left.  -N\left(
\hbar\omega_{nk\sigma}-\mu_{nk\sigma}^{\left(  2\right)  },T^{\left(
2\right)  }\right)  \right]  .
\end{align}
In the ballistic limit, $t_{n^{\prime}\sigma^{\prime}n\sigma}=\delta
_{nn^{\prime}}\delta_{\sigma\sigma^{\prime}}$
\begin{align}
J_{q} &  =-\sum_{n\sigma}\int_{\omega_{n\sigma}^{\mathrm{min}}}^{\omega
_{n\sigma}^{\mathrm{max}}}\frac{d\omega}{2\pi}\hbar\omega\left[  N\left(
\hbar\omega-\mu_{n\sigma}^{\left(  2\right)  }\left(  \omega\right)
,T^{\left(  2\right)  }\right)  \right.  \nonumber\\
&  \left.  -N\left(  \hbar\omega-\mu_{n\sigma}^{\left(  1\right)  }\left(
\omega\right)  ,T^{\left(  1\right)  }\right)  \right]  ,
\end{align}
where $\omega_{n\sigma}^{\mathrm{min}}$ and $\omega_{n\sigma}^{\mathrm{max}}$
are the band edges of the $n\sigma$ phonon mode, which in the absence of an
electric field or ferroelectric order reduces to the conventional phonon heat
current expression \cite{rego1998quantized,blencowe1999quantum}. We obtain the
transport coefficients in Eq.~(\ref{resp}) by linearizing the distribution
functions:
\begin{align}
G &  =-\frac{1}{\hbar}\sum_{n\sigma}\int_{\omega_{n\sigma}^{\mathrm{min}}%
}^{\omega_{n\sigma}^{\mathrm{max}}}\xi_{n\sigma}^{2}\left(  \omega\right)
\left(  \frac{\partial N}{\partial\omega}\right)  _{T}\frac{d\omega}{2\pi
},\nonumber\\
\Pi &  =ST=G^{-1}\sum_{n\sigma}\int_{\omega_{n\sigma}^{\mathrm{min}}}%
^{\omega_{n\sigma}^{\mathrm{max}}}\omega\xi_{n\sigma}\left(  \omega\right)
\left(  \frac{\partial N}{\partial\omega}\right)  _{T}\frac{d\omega}{2\pi
},\nonumber\\
K &  =\sum_{n\sigma}\int_{\omega_{n\sigma}^{\mathrm{min}}}^{\omega_{n\sigma
}^{\mathrm{max}}}\hbar\omega\left(  \frac{\partial N}{\partial T}\right)
_{\omega}\frac{d\omega}{2\pi},\label{Trans}%
\end{align}
which (for simple ferroelectrics) are positive since $\partial N(\omega
,T)/\partial\omega<0$. \begin{figure*}[ptb]
\centering
\par
\includegraphics[width=14.0 cm]{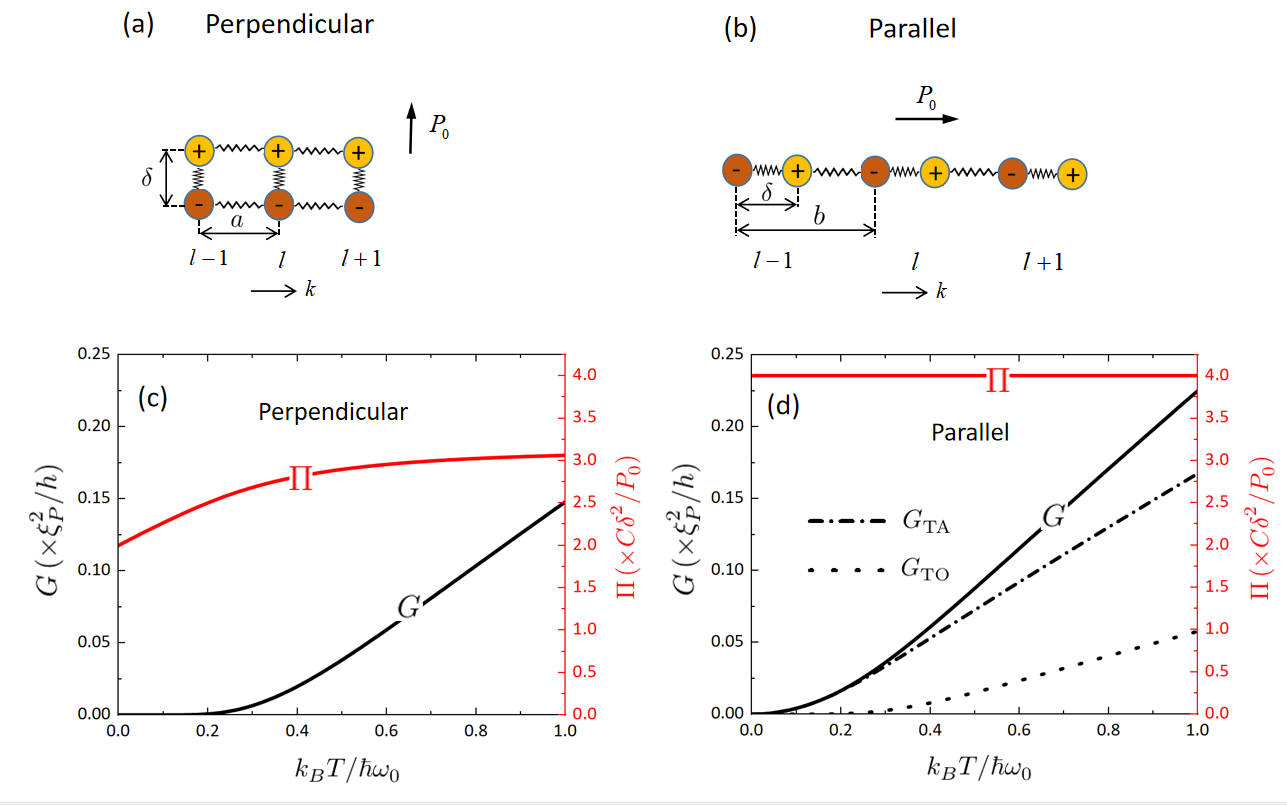}\newline\caption{Diatomic
ferroelectric chains with spontaneous polarization (a) perpendicular and (b)
parallel to the chain axis, respectively. The polarization conductance ($G$)
and Peltier coefficient ($\Pi$) as a function of temperature for the
perpendicular (c) and parallel (d) configurations, where TA and TO in (d)
represent the contributions from the transverse acoustic and optical phonons,
respectively.}%
\label{Fig-2}%
\end{figure*}

\emph{Diatomic ferroelectric chain:} For concreteness, we model transport
properties at temperatures sufficiently below the ordering transition by a
one-dimensional dimer chain, with two ions of opposite charges $\pm Q$ and
same mass $M$ in the unit cell, as sketched in Fig.~\ref{Fig-2} for the
polarization perpendicular and parallel to the chain. The ground state
permanent electric dipole in each unit cell is $P_{0}=Q\delta$, where $\delta$
is the symmetry breaking deformation and $Q$ the ionic charge. For strongly
anisotropic systems this model holds by multiplying the results with the
number of parallel wires.

In the perpendicular configuration, only two optical phonons with polarization
vector transverse to the ferroelectric order carry an average polarization of
$\xi_{k\sigma}=-(\hbar Q^{2})/(MP_{0}\omega_{k\sigma})$. Using the results
from the Supplementary Material (SM) \cite{SM}
\begin{align}
G  &  =\frac{2\xi_{P}^{2}}{h}\int_{\sqrt{2}\epsilon_{0}}^{\sqrt{6}\epsilon
_{0}}d\epsilon\frac{\epsilon_{0}^{2}e^{\epsilon}}{\epsilon^{2}(e^{\epsilon
}-1)^{2}}\nonumber\\
&  =\frac{\xi_{P}^{2}}{h}\left\{
\begin{array}
[c]{c}%
e^{-\sqrt{2}\epsilon_{0}}-\frac{1}{3}e^{-\sqrt{6}\epsilon_{0}}\\
\left(  \frac{1}{3\sqrt{2}}-\frac{1}{9\sqrt{6}}\right)  \frac{k_{B}T}%
{\hbar\omega_{0}}%
\end{array}
\text{ for}%
\begin{array}
[c]{c}%
k_{B}T\ll\hbar\omega_{0}\\
k_{B}T\gg\hbar\omega_{0}%
\end{array}
\right.  ,
\end{align}%
\begin{align}
\Pi &  =\left(  \frac{2\xi_{P}^{2}}{hG}\right)  \frac{C\delta^{2}}{P_{0}}%
\int_{\sqrt{2}\epsilon_{0}}^{\sqrt{6}\epsilon_{0}}d\epsilon\frac{e^{\epsilon}%
}{(e^{\epsilon}-1)^{2}}\nonumber\\
&  =\frac{C\delta^{2}}{P_{0}}\left\{
\begin{array}
[c]{c}%
2\\
\frac{18}{13}(4-\sqrt{3})
\end{array}
\text{ for }%
\begin{array}
[c]{c}%
k_{B}T\ll\hbar\omega_{0}\\
k_{B}T\gg\hbar\omega_{0}%
\end{array}
\right.  ,
\end{align}%
\begin{align}
K  &  \approx\frac{k_{B}^{2}T}{h}\left[  \int_{0}^{2\epsilon_{0}}%
d\epsilon\frac{3\epsilon^{2}e^{\epsilon}}{(e^{\epsilon}-1)^{2}}+\int_{\sqrt
{2}\epsilon_{0}}^{\sqrt{6}\epsilon_{0}}d\epsilon\frac{2\epsilon^{2}%
e^{\epsilon}}{(e^{\epsilon}-1)^{2}}\right] \nonumber\\
&  =\left\{
\begin{array}
[c]{c}%
\frac{\pi^{2}k_{B}^{2}T}{h}\\
\left(  3+\sqrt{6}-\sqrt{2}\right)  \frac{k_{B}\omega_{0}}{\pi}%
\end{array}
\text{ for }%
\begin{array}
[c]{c}%
k_{B}T\ll\hbar\omega_{0}\\
k_{B}T\gg\hbar\omega_{0}%
\end{array}
\right.  ,
\end{align}
where $\omega_{0}=\sqrt{C/M}$ is the characteristic frequency of lattice
vibration with $C$ the shear force constant, $\xi_{P}=\hbar\omega_{0}%
P_{0}/(C\delta^{2})$ is a quantum polarization, analogous to the Bohr magneton
for magnetization with $-\xi_{P}E/\hbar$ equivalent to the Rabi frequency,
$\epsilon_{0}=\hbar\omega_{0}/k_{B}T$ and $(0,2\epsilon_{0})$ and $(\sqrt
{2}\epsilon_{0},\sqrt{6}\omega_{0})$ are the band edges in unit of $k_{B}T$
for three acoustic and two transverse optical phonon modes, respectively
\cite{SM}. The ratio between the elastic energy and electric dipole
$C\delta^{2}/P_{0}$ also governs transport in the diffuse model
\cite{PhysRevLett.126.187603}. In Fig.~\ref{Fig-2}(c), we plot the
polarization conductance and Peltier coefficient as a function of temperature
(see Fig. S2 in SM \cite{SM} for the heat conductance). Since the polarization
transport is contributed by gapped optical phonons, $G$ is exponentially small
when $k_{B}T\ll\hbar\omega_{0}$. In contrast to $K=\pi^{2}k_{B}^{2}T/h$, the
well-known quantum of phononic heat conductance
\cite{schwab2000measurement,rego1998quantized,blencowe1999quantum}, the
polarization Peltier quantum $\Pi=C\delta^{2}/P_{0}$ is not universal, but
depends on the material parameters.

Aligning the polarization with the wire axis drastically changes the
polarization transport that is then carried by both transverse acoustic and
optical phonon modes. According to the SM \cite{SM}
\begin{equation}
\xi_{k\sigma}=-\frac{\hbar\omega_{k\sigma}P_{0}}{4C\delta^{2}},
\end{equation}
where $\sigma=\mathrm{TA},\mathrm{TO}$ denotes the transverse acoustic and
optical phonons, respectively, and Eq.~(\ref{Trans}) reduces to
\begin{equation}
K=\frac{k_{B}^{2}T}{h}\int_{0}^{2\epsilon_{0}}d\epsilon\frac{3\epsilon
^{2}e^{\epsilon}}{(e^{\epsilon}-1)^{2}}=\left\{
\begin{array}
[c]{c}%
\frac{\pi^{2}k_{B}^{2}T}{h}\\
\frac{3k_{B}\omega_{0}}{\pi}%
\end{array}
\text{for }%
\begin{array}
[c]{c}%
k_{B}T\ll\hbar\omega_{0}\\
k_{B}T\gg\hbar\omega_{0}%
\end{array}
\right.  ,
\end{equation}%
\begin{equation}
G=\frac{\xi_{P}^{2}}{8h}\int_{0}^{2\epsilon_{0}}d\epsilon\frac{\epsilon
^{2}e^{\epsilon}}{\epsilon_{0}^{2}(e^{\epsilon}-1)^{2}}=\frac{2}{3}\frac
{KT}{\Pi^{2}},
\end{equation}%
\begin{equation}
\Pi=\frac{4C\delta^{2}}{P_{0}}.
\end{equation}
Here $\Pi$ is constant and the conductance $G$ vanishes quadratically with
temperature since polarization transport by the transverse acoustic phonons is
massless at low energies. The figure of merit of thermal polarization
transport turns out to be constant as well:
\begin{equation}
ZT\equiv\frac{G\Pi^{2}}{KT}=\frac{2}{3}%
\end{equation}

\emph{Detection of polarization current:} In the steady state the polarization
current from the high-field to the low field region is accompanied by a heat
current. We assume in the derivations above that the reservoirs are such large
that on the time scale of the transport process the bias is constant. Finite
reservoirs react parametrically to these currents on a larger time scale. The
Peltier effect can be observed by an increased temperature in the high-field
regime and cooling of the low-field terminal, while the \textquotedblleft
battery\textquotedblright\ becomes depleted. When the two reservoirs are not
electrically biased but subject to a temperature difference, a heat current
flows, accompanied by a Seebeck polarization current. Both currents contribute
to a increase (decrease) of the polarization on the hot (cold) side that
charges the capacitors by the pyroelectric as well as Seebeck effect, i.e.
generates thermovoltages in both reservoirs. With the parameters above we can
compute the time dependence for given sample geometries parametrically.\ 

The dc transport of electric polarization is accompanied by dc stray magnetic
fields and, vice versa, an applied magnetic field can affect the polarization
current. When flowing along the $x$ with polarization along $z$ as in
Fig.~1(a), the magnetic flux density at a position $\mathbf{r}=(0,y,z)$
follows from the Lorentz transformation
\begin{equation}
\mathbf{B}_{p}=\frac{\mu_{0}J_{p}}{2\pi\rho^{2}}\left(  0,\cos2\phi,\sin
2\phi\right)  \label{MF}%
\end{equation}
where $J_{p}$ is the polarization current, $\rho=\sqrt{y^{2}+z^{2}}$ the
probing distance, $\cos\phi=y/\rho$ and $\sin\phi=z/\rho$. For $N_{0}$
uncoupled parallel chains $J_{p}=N_{0}G[\Delta E+\Pi(\Delta T/T)]$, where $G$
and $\Pi$ are the polarization conductance and Peltier coefficient for the
single chain, respectively. In the perpendicular polarization configuration at
$T=300\,\mathrm{K}$, with $C=25\,\mathrm{J/m^{2}}$, $\omega_{0}%
=10\,\mathrm{THz}$, $\delta=0.03\,\mathrm{nm}$, $P_{0}=2\times10^{-29}%
\,\mathrm{Cm}$, we arrive at $G=9.74\times10^{-28}\,\mathrm{m^{2}/\Omega}$,
$\Pi=3.52\times10^{9}\,\mathrm{V/m.}$ The induced magnetic field for $N_{0}=1$
at a distance $\rho=10\,\mathrm{nm}$ is $B_{p}\approx200\,\mathrm{pT}$ for
either $\triangle E=10^{8}\,\mathrm{V/m}$ or $\Delta T=10\,\mathrm{K}$, which
can be detected by single diamond-NV center magnetometry enhanced by
spin-to-charge NV readout protocols \cite{Andersen}. The stray magnetic fields
generated by the polarization current in thicker wires and tuned by the sample
geometry may become large enough to be detectable by conventional sensors.

\emph{Conclusions:} We derive expressions for the steady state polarization
and heat transport through a ferroelectric constriction driven by temperature
and electric field differences. We find drastic effects of rotating the
polarization direction, such as an algebraic vs. exponential suppression of
the polarization current. The results can be extended to include, e.g., the
effects of ferroelectric domain walls. The polarization current can be
detected indirectly via the polarization Peltier effect and a thermovoltage
or, more directly, by the stray magnetic field that accompanies the streaming
dipoles. Our formulation for the polarization transport is not limited to this
simple chain model but is accessible to first-principles calculations. The
thermally and electrically induced transport of electric polarization opens
alternative strategies for thermal management using ferroelectric materials.

\emph{Acknowledges:} We acknowledge helpful discussions with Toeno van der
Sar, Tomoya Nakatani, Toshu An, Kei Yamamoto and Weichao Yu. This work is
supported by JSPS KAKENHI Grant No. 19H006450.
\textit{ }



\begin{thebibliography}{0}%
\makeatletter
\providecommand \@ifxundefined [1]{%
 \@ifx{#1\undefined}
}%
\providecommand \@ifnum [1]{%
 \ifnum #1\expandafter \@firstoftwo
 \else \expandafter \@secondoftwo
 \fi
}%
\providecommand \@ifx [1]{%
 \ifx #1\expandafter \@firstoftwo
 \else \expandafter \@secondoftwo
 \fi
}%
\providecommand \natexlab [1]{#1}%
\providecommand \enquote  [1]{``#1''}%
\providecommand \bibnamefont  [1]{#1}%
\providecommand \bibfnamefont [1]{#1}%
\providecommand \citenamefont [1]{#1}%
\providecommand \href@noop [0]{\@secondoftwo}%
\providecommand \href [0]{\begingroup \@sanitize@url \@href}%
\providecommand \@href[1]{\@@startlink{#1}\@@href}%
\providecommand \@@href[1]{\endgroup#1\@@endlink}%
\providecommand \@sanitize@url [0]{\catcode `\\12\catcode `\$12\catcode
  `\&12\catcode `\#12\catcode `\^12\catcode `\_12\catcode `\%12\relax}%
\providecommand \@@startlink[1]{}%
\providecommand \@@endlink[0]{}%
\providecommand \url  [0]{\begingroup\@sanitize@url \@url }%
\providecommand \@url [1]{\endgroup\@href {#1}{\urlprefix }}%
\providecommand \urlprefix  [0]{URL }%
\providecommand \Eprint [0]{\href }%
\providecommand \doibase [0]{http://dx.doi.org/}%
\providecommand \selectlanguage [0]{\@gobble}%
\providecommand \bibinfo  [0]{\@secondoftwo}%
\providecommand \bibfield  [0]{\@secondoftwo}%
\providecommand \translation [1]{[#1]}%
\providecommand \BibitemOpen [0]{}%
\providecommand \bibitemStop [0]{}%
\providecommand \bibitemNoStop [0]{.\EOS\space}%
\providecommand \EOS [0]{\spacefactor3000\relax}%
\providecommand \BibitemShut  [1]{\csname bibitem#1\endcsname}%
\let\auto@bib@innerbib\@empty
\end{thebibliography}%


\begin{thebibliography}{99}                                                                                               %


\bibitem {maxwell1873treatise}J.~C. Maxwell,
\newblock {\em A treatise on electricity and magnetism}, Volume~1 (Oxford:
Clarendon Press, 1873).

\bibitem {sharvin1965possible}Y.~V. Sharvin,
\newblock {Zh. Eksperim. i Teor. Fiz.} \textbf{48} (1965).

\bibitem {van1992thermo}H. van~Houten, L.~W. Molenkamp, C.~W.~J. Beenakker,
and C.~T. Foxon,
\newblock {Semiconductor Science and Technology} \textbf{7}, B215 (1992).

\bibitem {van1996quantum}H. van~Houten and C.~W.~J. Beenakker,
\newblock {Physics Today} \textbf{49}, 22 (1996).

\bibitem {agrait2003quantum}N.~Agra{\i}t, A.~L. Yeyati, and J.~M.
Van~Ruitenbeek,
\newblock {Physics Reports} \textbf{377}, 81 (2003).

\bibitem {van1988quantized}B.~J. Van~Wees, H.~Van~Houten, C.~W.~J. Beenakker,
J.~G. Williamson, L.~P. Kouwenhoven, D.~Van~der Marel, and C.~T. Foxon,
\newblock {Physical Review Letters} \textbf{60}, 848 (1988).

\bibitem {wharam1988one}D.~A. Wharam, T.~J. Thornton, R.~Newbury, M.~Pepper,
H.~Ahmed, J.~E.~F. Frost, D.~G. Hasko, D.~C. Peacock, D.~A. Ritchie, and
G.~A.~C Jones,
\newblock {Journal of Physics C: Solid State Physics} \textbf{21}, L209 (1988).

\bibitem {montie1991observation}E.~A. Montie, E.~C. Cosman, G.~{W.'t} Hooft,
M.~B. Van~der Mark, and C.~W.~J. Beenakker,
\newblock {Nature} \textbf{350}, 594 (1991).

\bibitem {beenakker1991josephson}C.~W.~J. Beenakker and H.~Van~Houten,
\newblock {Physical Review Letters} \textbf{66}, 3056 (1991).

\bibitem {takayanagi1995observation}H.~Takayanagi, T.~Akazaki, and J.~Nitta,
\newblock {Physical Review Letters} \textbf{75}, 3533 (1995).

\bibitem {debray2009all}P.~Debray, S.~M.~S. Rahman, J.~Wan, R.~S. Newrock,
M.~Cahay, A.~T. Ngo, S.~E. Ulloa, S.~T. Herbert, M.~Muhammad, and M.~Johnson,
\newblock {Nature Nanotechnology} \textbf{4}, 759 (2009).

\bibitem {meier2003magnetization}F.~Meier and D.~Loss,
\newblock {Physical Review Letters} \textbf{90}, 167204 (2003).

\bibitem {meschke2006single}M.~Meschke, W.~Guichard, and J.~P. Pekola,
\newblock {Nature} \textbf{444}, 187 (2006).

\bibitem {ojanen2007photon}T.~Ojanen and T.~T. Heikkil{\"a},
\newblock {Physical Review B} \textbf{76}, 073414 (2007).

\bibitem {chiatti2006quantum}O.~Chiatti, J.~T. Nicholls, Y.~Y. Proskuryakov,
N.~Lumpkin, I.~Farrer, and D.~A. Ritchie,
\newblock {Physical Review Letters} \textbf{97}, 056601 (2006).

\bibitem {jezouin2013quantum}S.~Jezouin, F.~D. Parmentier, A.~Anthore,
U.~Gennser, A.~Cavanna, Y.~Jin, and F.~Pierre,
\newblock {Science} \textbf{342}, 601 (2013).

\bibitem {cui2017quantized}L.~Cui, W.~Jeong, S.~Hur, M.~Matt, J.~C.
Kl{\"o}ckner, F.~Pauly, P.~Nielaba, J.~C. Cuevas, E.~Meyhofer, and P.~Reddy,
\newblock {Science} \textbf{355}, 1192 (2017).

\bibitem {rego1998quantized}L.~G.~C. Rego and G.~Kirczenow,
\newblock {Physical Review Letters} \textbf{81}, 232 (1998).

\bibitem {blencowe1999quantum}M.~P. Blencowe,
\newblock {Physical Review B} \textbf{59}, 4992 (1999).

\bibitem {schwab2000measurement}K.~Schwab, E.~A. Henriksen, J.~M. Worlock, and
M.~L. Roukes,
\newblock {Nature} \textbf{404}, 974 (2000).

\bibitem {xu2013ferroelectric}Y.~H. Xu,
\newblock {\em Ferroelectric materials and their applications} (Elsevier, 2013).

\bibitem {spaldin2007analogies}N.~A. Spaldin,
\newblock {Physics of ferroelectrics}, Pages 175--218 (2007).

\bibitem {kajiwara2010transmission}Y.~Kajiwara, K.~Harii, S.~Takahashi, J.
Ohe, K.~Uchida, M.~Mizuguchi, H.~Umezawa, H.~Kawai, K.~Ando, K.~Takanashi,
S.~Maekawa, and E.~Saitoh,
\newblock {Nature} \textbf{464}, 262 (2010).

\bibitem {cornelissen2016magnon}L.~J. Cornelissen, K.~J.~H. Peters, G.~E.~W.
Bauer, R.~A. Duine, and B.~J. van Wees,
\newblock {Physical Review B} \textbf{94}, 014412 (2016).

\bibitem {uchida2010spin}K. Uchida, J.~Xiao, H.~Adachi, Jun-ichiro Ohe,
S.~Takahashi, J.~Ieda, T.~Ota, Y.~Kajiwara, H.~Umezawa, H.~Kawai, G.~E.~W
Bauer, S.~Maekawa, and E.~Saitoh,
\newblock {Nature Materials} \textbf{9}, 894 (2010).

\bibitem {flipse2014observation}J.~Flipse, F.~K. Dejene, D.~Wagenaar, G.~E.~W.
Bauer, J.~B. Youssef, and B.~J. Van~Wees,
\newblock {Physical Review Letters} \textbf{113}, 027601 (2014).

\bibitem {daimon2016thermal}S.~Daimon, R.~Iguchi, T.~Hioki, E.~Saitoh, and .
Uchida,
\newblock {Nature Communications} \textbf{7}, 1 (2016).

\bibitem {PhysRevLett.126.187603}G. E.~W. Bauer, R. Iguchi, and K. Uchida,
\newblock {Physical Review Letter} \textbf{126}, 187603 (2021).

\bibitem {buttiker1992scattering}M.~B{\"u}ttiker,
\newblock {Physical Review B} \textbf{46}, 12485 (1992).

\bibitem {datta1997electronic}S.~Datta, \emph{Electronic transport in
mesoscopic systems} (Cambridge University Press, 1997).

\bibitem {born1954dynamical}M.~Born and K.~Huang,
\newblock {\em Dynamical theory of crystal lattices} (Clarendon Press, 1954).

\bibitem {freire1988lattice}J.~D. Freire and R.~S. Katiyar,
\newblock {Physical Review B} \textbf{37}, 2074 (1988).

\bibitem {pouget1986lattice}J.~Pouget, A.~A{\c{s}}kar, and G.~A. Maugin,
\newblock {Physical Review B} \textbf{33}, 6304 (1986).

\bibitem {SM}See the Supplementary Material for technical details.

\bibitem {Andersen}T.I. Andersen, B.L. Dwyer, J. D. Sanchez-Yamagishi, J.F.
Rodriguez-Nieva, K. Agarwal, K. Watanabe, T. Taniguchi, E.A. Demler, P. Kim,
H. Park, and M.D. Lukin, Science \textbf{364}, 154 (2019).
\end{thebibliography}
\end{document}


\title{Supplementary Material to \textquotedblleft The Ferroelectric Point
Contact\textquotedblright}
\author{Ping Tang$^{1}$}
\author{Ryo Iguchi$^{2}$}
\author{Ken-ichi Uchida$^{2,3,4}$}
\author{Gerrit E. W. Bauer$^{1,3,4,5}$}
\affiliation{$^1$WPI-AIMR, Tohoku
University, 2-1-1 Katahira, 980-8577 Sendai, Japan}
\affiliation{$^2$National Institute for Materials Science, Tsukuba 305-0047, Japan}
\affiliation{$^3$Institute for Materials Research, Tohoku University, 2-1-1 Katahira, 980-8577 Sendai, Japan}
\affiliation{$^4$Center for Spintronics Research Network, Tohoku University, Sendai 980-8577, Japan}
\affiliation{$^5$Zernike Institute for Advanced Materials, University of Groningen, 9747 AG Groningen, Netherlands}
\maketitle

\begin{figure}[h]
\centering
\par
\includegraphics[width=14.2 cm]{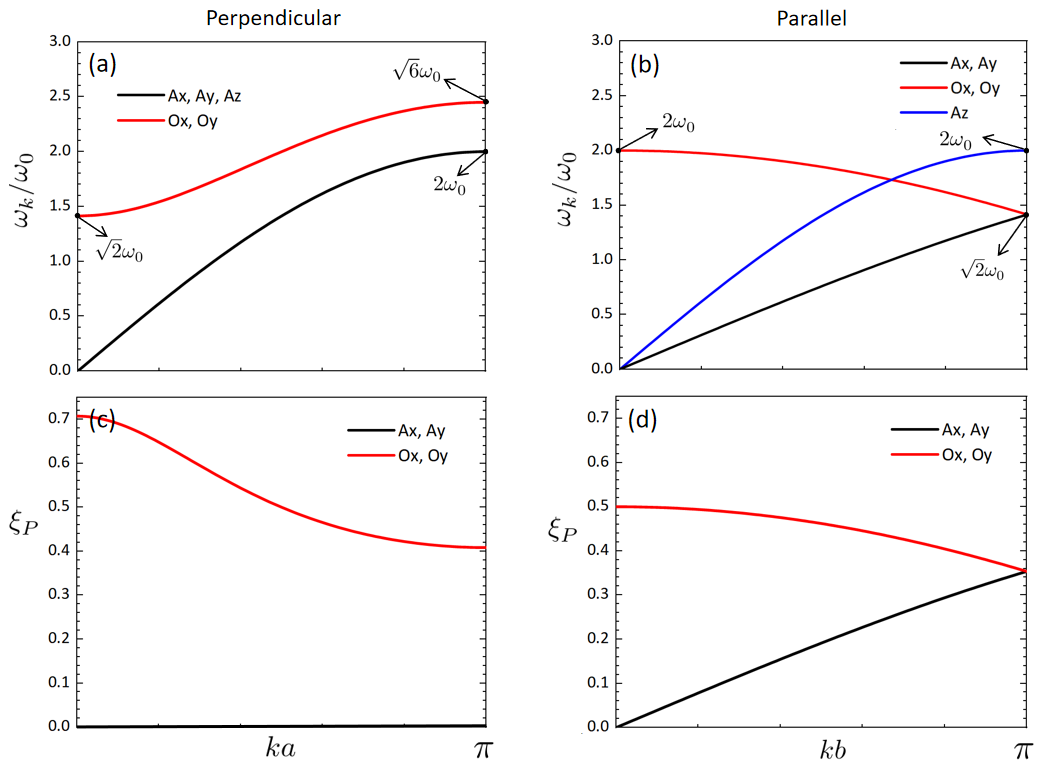}\newline\caption{The phonon band
structures for one-dimensional diatomic ferroelectric chains with ground state
polarization (a) perpendicular and (b) parallel to the chain axis,
respectively, where the direction of ground state polarization is fixed along
$z$-direction. In (c) and (d) we plot the polarization carried by each phonon
mode corresponding to (a) and (b), respectively. In the perpendicular
configuration, only the transverse optical modes carry polarization, while in
the parallel configuration the transverse acoustic modes also contribute. Here
the parameters are taken as $C_{0}^{(x)}=C_{0}^{(y)}=C_{1}^{(x)}=C_{1}%
^{(y)}=C_{1}^{(z)}=C$ and $C_{0}^{(z)}/C\gg1$ for (a) and $C_{0}^{(x)}%
=C_{0}^{(y)}=C_{1}^{(x)}=C_{1}^{(y)}=C,\,C_{1}^{(z)}=2\,C$ and $C_{0}%
^{(z)}/C\gg1$ for (b). The high-frequency longitudinal optical mode ($Oz$)
lies outside the plotting range of both (a) and (b).}%
\label{Fig-SM1}%
\end{figure}

\begin{figure}[h]
\centering
\par
\includegraphics[width=14.2 cm]{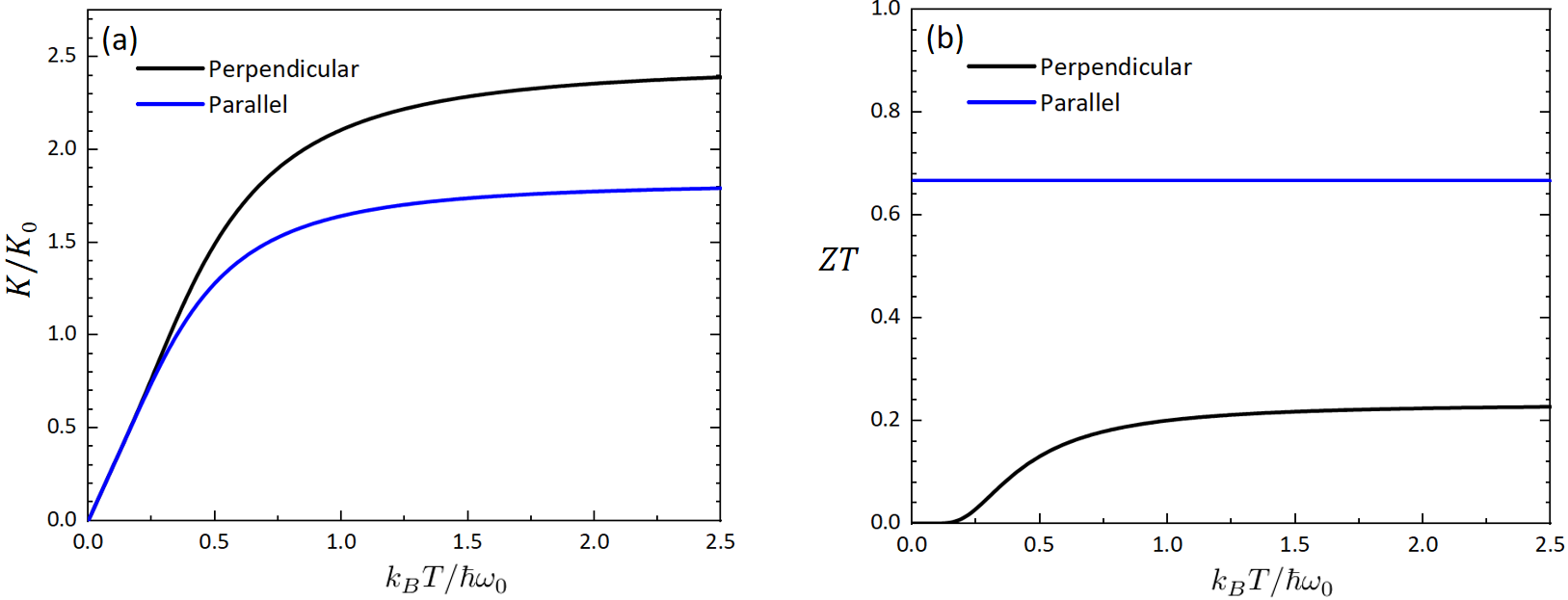}\newline\caption{(a) The heat
conductance ($K$) and (b) figure of merit ($ZT\equiv G\Pi^{2}/KT$) as a
function of temperature for ferroelectric order perpendicular and parallel to
the chain. $K$ is in units of $K_{0}=\pi k_{B}\omega_{0}/6$.}%
\label{Fig-SM2}%
\end{figure}

In this Supplementary Material, we derive the phonon dispersions and transport
coefficients for the configurations with ferroelectric polarization
perpendicular (Section A) and parallel (Section B) to the chain axis as used
in the main text \cite{PT}. Fig. \ref{Fig-SM1} summarized the results for the
phonon band structure. \ref{Fig-SM2} on the thermal conductance and figure of
merit supplement the figures given in the main text.    

\section{A. Perpendicular polarization configuration}

The energy in a displacive ferroelectric has both mechanical and electrostatic
contributions. The elastic energy cost of small deformations from equilibrium
in our chain model with a perpendicular configuration is
\begin{equation}
\Phi_{\mathrm{el}}=\frac{1}{2}\sum_{l\alpha}C_{0}^{(\alpha)}[u_{l}^{(\alpha
)}-v_{l}^{(\alpha)}]^{2}+\frac{1}{2}\sum_{l\alpha}C_{1}^{\alpha}%
[u_{l}^{(\alpha)}-u_{l+1}^{(\alpha)}]^{2}+\frac{1}{2}\sum_{l\alpha}%
C_{1}^{(\alpha)}[v_{l}^{(\alpha)}-v_{l+1}^{(\alpha)}]^{2},\nonumber
\end{equation}
where $\alpha=x,y,z$ denotes the Cartesian components and $u_{l}^{(\alpha)}$
and $v_{l}^{(\alpha)}$ the displacement of positive and negative ions of unit
cell $l$ in the $\alpha$-direction, respectively. $C_{0}^{(\alpha)}$ is the
corresponding $\alpha$-th component force constant between the positive and
negative ions in the unit cell and $C_{1}^{(\alpha)}$ that between the same
type of ions in neighboring unit cells, for simplicity assumed to be the same
for the positive and negative ions.

In a ferroelectric, we should also consider the depolarization cost of the
dipolar energy
\begin{equation}
\Phi_{\mathrm{dp}}=\frac{1}{8\pi\varepsilon}\sum_{l\neq l^{\prime}}%
\frac{\mathbf{P}_{l}\cdot\mathbf{P}_{l^{\prime}}-3(\mathbf{P}_{l}\cdot
\hat{\mathbf{r}}_{ll^{\prime}})(\mathbf{P}_{l^{\prime}}\cdot\hat{\mathbf{r}%
}_{ll^{\prime}})}{R_{ll^{\prime}}^{3}},\tag{$\rm S.A1$}\label{dipol}%
\end{equation}
where $\varepsilon$ the dielectric constant, $R_{ll^{\prime}}$ the distance
between $l$-th and $l^{\prime}$-th unit cell with direction vector
$\hat{\mathbf{r}}_{ll^{\prime}}$ and $\mathbf{P}_{l}=[1-(\delta\mathbf{P}%
_{l}^{\perp})^{2}/(2P_{0}^{2})]\mathbf{P}_{0}+\delta\mathbf{P}_{l}$ the
polarization of each unit cell with $\mathbf{P}_{0}$ the polarization at the
ground state and $\delta\mathbf{P}_{l}=Q(\mathbf{u}_{l}-\mathbf{v}_{l})$ the
polarization fluctuations \cite{PT}. The dipolar interaction contains terms
that are linear in the ion displacement and affect the equilibrium position of
the atoms. These can be removed by defining a new ground state and
subsequently disregarded.

With chain axis along $x$ and $\mathbf{P}_{0}$ along $z$ direction, a plane
wave ansatz with wave number $k$ leads to a dynamical matrix in the $\alpha
$-direction
\begin{align}
D^{(\alpha)}(k,s,s^{\prime})  &  \equiv\frac{1}{M}\sum_{l^{\prime}}\left(
\begin{matrix}
\frac{\partial^{2}\Phi}{\partial u_{l}^{(\alpha)}\partial u_{l^{\prime}%
}^{(\alpha)}}, & \frac{\partial^{2}\Phi}{\partial u_{l}^{(\alpha)}\partial
v_{l^{\prime}}^{(\alpha)}}\\
\frac{\partial^{2}\Phi}{\partial v_{l}^{(\alpha)}\partial u_{l^{\prime}%
}^{(\alpha)}}, & \frac{\partial^{2}\Phi}{\partial v_{l}^{(\alpha)}\partial
v_{l^{\prime}}^{(\alpha)}}%
\end{matrix}
\right)  _{ss^{\prime}}e^{ik(l-l^{\prime})a}\nonumber\\
&  =\frac{1}{M}\left(
\begin{matrix}
2C_{1}^{(\alpha)}(1-\cos ka)+\widetilde{C}_{0}^{(\alpha)}+C_{d}^{(\alpha
)}(k), & -\widetilde{C}_{0}^{(\alpha)}-C_{d}^{(\alpha)}(k)\\
-\widetilde{C}_{0}^{(\alpha)}-C_{d}^{(\alpha)}(k), & 2C_{1}^{(\alpha)}(1-\cos
ka)+\widetilde{C}_{0}^{(\alpha)}+C_{d}^{(\alpha)}(k)
\end{matrix}
\right)  _{ss^{\prime}} \tag{$\rm S.A2$}\label{dyn}%
\end{align}
where $\Phi=\Phi_{\mathrm{el}}+\Phi_{\mathrm{dp}}$ is the total energy cost by
the lattice deformation, $s,s^{\prime}=\pm$ denotes the type of ions, $M$ the
ionic mass and $a$ the lattice constant; $\widetilde{C}_{0}^{(\alpha)}%
=C_{0}^{(\alpha)}[1-2\zeta(3)\eta^{(\alpha)}(\delta_{\alpha,x}+\delta
_{\alpha,y})]$ is the dipolar-interaction modified force constant for the ions
in the same unit cell, and
\begin{equation}
C_{d}^{(\alpha)}(k)=2(1-3\delta_{\alpha,x})\eta^{(\alpha)}\left[  \sum
_{n=1}^{\infty}\cos(nka)/n^{3}\right]  C_{0}^{(\alpha)} \nonumber
\end{equation}
is generated by the dipolar interaction and $\eta^{(\alpha)}=Q^{2}%
/(4\pi\varepsilon a^{3}C_{0}^{(\alpha)})$ is a dimensionless constant
representing the relative strength of the dipolar interaction and $\sum
_{n=1}^{\infty}n^{-3}=\zeta\left(  3\right)  $. For $a=0.4\,$nm, $Q=e$,
$C_{0}^{(\alpha)}=25\,$J/m$^{2}$ and a dielectric constant $\varepsilon
/\varepsilon_{0}=10^{3}$ in a typical perovskite ferroelectric, $\eta
^{(\alpha)}\sim10^{-4}$, where $\varepsilon_{0}$ is the vacuum dielectric
constant. We therefore may discard the effect of the dipolar interaction on
the phonon dispersions. We note that in ferromagnets the exchange interaction
vanishes for low-frequency spin waves like $\left(  ka\right)  ^{2}$ while the
dipolar scales like $ka.$ In ferroelectrics the leading elastic energy is a
large constant for the electrically active optical modes.

The eigenvalues and eigenvectors of the dynamical matrix are the phonon
dispersion relation and polarization vectors, respectively,
\begin{align}
\omega_{kO}^{(\alpha)}  &  =\sqrt{\frac{2C_{1}^{(\alpha)}}{M}(1-\cos
ka)+\frac{2C_{0}^{(\alpha)}}{M}},\;\mathbf{e}_{kO}^{(\alpha)}=\frac{1}%
{\sqrt{2}}\left(  1,-1\right)  ,\nonumber\\
\omega_{kA}^{(\alpha)}  &  =\sqrt{\frac{2C_{1}^{(\alpha)}}{M}(1-\cos
ka)},\;\mathbf{e}_{kA}^{(\alpha)}=\frac{1}{\sqrt{2}}\left(  1,1\right)  ,
\tag{$\rm S.A3$}\label{Disp}%
\end{align}
where $O$ and $A$ are the optical and acoustic modes, respectively. The two
components of $\mathbf{e}_{kO}^{(\alpha)}$ represent the relative phase of the
two ions in the unit cell (rather than the spatial direction $\alpha$). The
vector $\mathbf{F}_{k\sigma}^{(\alpha)}=\sum_{s}Q_{s}/\sqrt{M_{s}}%
\mathbf{e}_{k\sigma}^{(\alpha)}(s)$ (defined in the main text) determines the
contribution of phonon to the polarization dynamics with
\begin{equation}
|\mathbf{F}_{kO}^{(\alpha)}|^{2}=\frac{2Q^{2}}{M},\,\,\,\,\,|\mathbf{F}%
_{kA}^{(\alpha)}|^{2}=0. \tag{$\rm S.A4$}%
\end{equation}
Thus, the acoustic phonons do not contribute to the polarization dynamics. At
a given $k$ two optical phonons $\alpha=x,y$ with polarization vector
transverse to the ferroelectric order (along $z$-direction) carry a
polarization of \cite{PT}
\begin{equation}
\xi_{kO}^{(\alpha)}\equiv-\frac{\hbar}{2P_{0}}\frac{|\mathbf{F}_{kO}%
^{(\alpha)}|^{2}}{\omega_{kO}^{(\alpha)}}=-\frac{\hbar Q^{2}}{MP_{0}%
\omega_{kO}^{(\alpha)}}. \tag{$\rm S.A5$}\label{polph}%
\end{equation}

We now assume that all the components of force constants are the same except
for the longitudinal one between the positive and negative ions in the same
unit cell that is taken to be much larger, which implies that the modulus of
the polarization does not change, i.e., $C_{0}^{(x)}=C_{0}^{(y)}=C_{1}%
^{(x)}=C_{1}^{(y)}=C_{1}^{(z)}=C$ and $C_{0}^{z}/C\gg1$. Eq.~(\ref{Disp}) then
reduces to
\begin{align}
\omega_{kO}^{(x)} &  =\omega_{kO}^{(y)}=2\omega_{0}\sqrt{1-\frac{1}{2}\cos
ka},\nonumber\\
\omega_{kO}^{(z)} &  \approx\sqrt{\frac{2C_{0}^{z}}{C}}\omega_{0}\gg\omega
_{0},\nonumber\\
\omega_{kA}^{(x)} &  =\omega_{kA}^{(y)}=\omega_{kA}^{(z)}=\sqrt{2}\omega
_{0}\sqrt{1-\cos ka},\tag{$\rm S.A6$}\label{rdisp}%
\end{align}
where $\omega_{0}=\sqrt{C/M}$. Using Eq.~(\ref{polph}) and Eq.~(\ref{rdisp}),
we semi-analytically obtain the transport coefficients in the main text:%
\begin{align}
G &  =-\frac{1}{\hbar}\sum_{\alpha=x,y}\int_{\sqrt{2}\omega_{0}}^{\sqrt
{6}\omega_{0}}[\xi_{kO}^{(\alpha)}]^{2}\frac{\partial N(\omega_{kO}^{(\alpha
)},T)}{\partial\omega_{kO}^{(\alpha)}}\frac{d\omega_{kO}^{(\alpha)}}{2\pi
}=\frac{2\xi_{P}^{2}}{h}\int_{\sqrt{2}\epsilon_{0}}^{\sqrt{6}\epsilon_{0}%
}d\epsilon\frac{\epsilon_{0}^{2}e^{\epsilon}}{\epsilon^{2}(e^{\epsilon}%
-1)^{2}}\nonumber\\
\Pi &  =ST=G^{-1}\sum_{\alpha=x,y}\int_{\sqrt{2}\omega_{0}}^{\sqrt{6}%
\omega_{0}}\omega_{kO}^{(\alpha)}\xi_{kO}^{(\alpha)}\frac{\partial
N(\omega_{kO}^{(\alpha)},T)}{\partial\omega_{kO}^{(\alpha)}}\frac{d\omega
_{kO}^{(\alpha)}}{2\pi}=\frac{C\delta^{2}}{P_{0}}\left(  \frac{2\xi_{P}^{2}%
}{Gh}\right)  \int_{\sqrt{2}\epsilon_{0}}^{\sqrt{6}\epsilon_{0}}d\epsilon
\frac{e^{\epsilon}}{(e^{\epsilon}-1)^{2}}\nonumber\\
K &  \approx\sum_{\alpha=x,y,z}\int_{0}^{2\omega_{0}}\hbar\omega_{kA}%
^{(\alpha)}\frac{\partial N(\omega_{kA}^{(\alpha)},T)}{\partial T}%
\frac{d\omega_{kA}^{(\alpha)}}{2\pi}+\sum_{\alpha=x,y}\int_{\sqrt{2}\omega
_{0}}^{\sqrt{6}\omega_{0}}\hbar\omega_{kO}^{(\alpha)}\frac{\partial
N(\omega_{kO}^{(\alpha)},T)}{\partial T}\frac{d\omega_{kO}^{(\alpha)}}{2\pi
}\nonumber\\
&  =\frac{k_{B}^{2}T}{h}\left[  \int_{0}^{2\epsilon_{0}}d\epsilon
\frac{3\epsilon^{2}e^{\epsilon}}{(e^{\epsilon}-1)^{2}}+\int_{\sqrt{2}%
\epsilon_{0}}^{\sqrt{6}\epsilon_{0}}d\epsilon\frac{2\epsilon^{2}e^{\epsilon}%
}{(e^{\epsilon}-1)^{2}}\right]  \tag{$\rm S.A7$}%
\end{align}
where $\epsilon_{0}=\hbar\omega_{0}/k_{B}T$, $\xi_{P}\equiv\hbar\omega
_{0}P_{0}/(C\delta^{2})$ with $\delta=P_{0}/Q$ the ionic deformation,
$(0,2\epsilon_{0})$ and $(\sqrt{2}\epsilon_{0},\sqrt{6}\epsilon_{0})$ are the
band edges in unit of $k_{B}T$ for the three acoustic and two transverse
optical phonon modes, respectively. Assuming $\omega_{kO}^{(z)}\gg\omega
_{0}\forall k,$ we disregarded the contribution of the longitudinal optical
phonon band ($Oz$) to the heat conductance.

The polarization transport by the two transverse optical modes is gapped and
therefore exponentially suppressed at low temperatures.

\section{B. Parallel polarization configuration}

When the electric dipoles point along the chain axis, the deformation costs
the energy
\begin{equation}
\Phi=\frac{1}{2}\sum_{l\alpha}C_{0}^{(\alpha)}[u_{l}^{(\alpha)}-v_{l}%
^{(\alpha)}]^{2}+\frac{1}{2}\sum_{l\alpha}C_{1}^{(\alpha)}[u_{l}^{(\alpha
)}-v_{l+1}^{(\alpha)}]^{2}, \tag{$\rm S.B1$}%
\end{equation}
with notation as above. Here we discarded the dipolar interaction, as
discussed in the previous section. Without loss of generality, we set both the
electric polarization ordering and chain axis along the $z$-direction, and the
lattice dynamical matrix reads,
\begin{equation}
D^{(\alpha)}(k,s,s^{\prime})=\frac{1}{M}\left(
\begin{matrix}
C_{0}^{(\alpha)}+C_{1}^{(\alpha)}, & -C_{0}^{(\alpha)}-C_{1}^{(\alpha)}%
e^{ikb}\\
-C_{0}^{(\alpha)}-C_{1}^{(\alpha)}e^{-ikb}, & C_{0}^{(\alpha)}+C_{1}%
^{(\alpha)}%
\end{matrix}
\right)  _{ss^{\prime}} \tag{$\rm S.B2$}\label{pdy}%
\end{equation}
where the lattice constant $b=2a$ since the dipolar symmetry breaking doubles
the size of the unit cell. This implies that the phonon bands are folded back
with gaps at the Brillouin zone boundaries $k=\pi/b$. The eigenvalues and
eigenvectors of the dynamic matrix are now
\begin{align}
\lbrack\omega_{kO}^{(\alpha)}]^{2}  &  =\frac{C_{0}^{(\alpha)}+C_{1}%
^{(\alpha)}}{M}\left(  1+\sqrt{1-\frac{2C_{0}^{(\alpha)}C_{1}^{(\alpha)}%
}{(C_{0}^{(\alpha)}+C_{1}^{(\alpha)})^{2}}(1-\cos kb)}\right) \nonumber\\
\lbrack\omega_{kA}^{(\alpha)}]^{2}  &  =\frac{C_{0}^{(\alpha)}+C_{1}%
^{(\alpha)}}{M}\left(  1-\sqrt{1-\frac{2C_{0}^{(\alpha)}C_{1}^{(\alpha)}%
}{(C_{0}^{(\alpha)}+C_{1}^{(\alpha)})^{2}}(1-\cos kb)}\right)  \tag{$\rm
S.B3$}%
\end{align}
and
\begin{equation}
\mathbf{e}_{kO}^{(\alpha)}=\frac{1}{\sqrt{2}}\left(  \frac{C_{0}^{(\alpha
)}+C_{1}^{(\alpha)}e^{ikb}}{|C_{0}^{(\alpha)}+C_{1}^{(\alpha)}e^{ikb}%
|},-1\right)  ,\,\,\,\mathbf{e}_{kA}^{(\alpha)}=\frac{1}{\sqrt{2}}\left(
\frac{C_{0}^{(\alpha)}+C_{1}^{(\alpha)}e^{ikb}}{|C_{0}^{(\alpha)}%
+C_{1}^{(\alpha)}e^{ikb}|},1\right)  \tag{$\rm S.B4$}%
\end{equation}
where $O$ and $A$ indicate again the optical and acoustic modes, respectively.
The vector relevant to the polarization dynamics reads now,
\begin{equation}
\mathbf{F}_{kO}^{(\alpha)}=\frac{Q}{\sqrt{2M}}\left(  \frac{C_{0}^{(\alpha
)}+C_{1}^{(\alpha)}e^{ikb}}{|C_{0}^{(\alpha)}+C_{1}^{(\alpha)}e^{ikb}%
|}+1\right)  \mathbf{e}_{\alpha},\,\,\,\,\,\mathbf{F}_{kA}^{(\alpha)}=\frac
{Q}{\sqrt{2M}}\left(  \frac{C_{0}^{(\alpha)}+C_{1}^{(\alpha)}e^{ikb}}%
{|C_{0}^{(\alpha)}+C_{1}^{(\alpha)}e^{ikb}|}-1\right)  \mathbf{e}_{\alpha}
\tag{$\rm S.B5$}%
\end{equation}
where $\mathbf{e}_{\alpha}$ is the unit vector along $\alpha$-direction. In
the long-wavelength limit ($k\rightarrow0$), $|\mathbf{F}_{kO}^{(\alpha)}%
|^{2}\rightarrow2Q^{2}/M$ and $|\mathbf{F}_{kA}^{(\alpha)}|^{2}\rightarrow0$.
However, in contrast to the perpendicular configuration, the transverse
acoustic phonons with finite $k$ now carry polarization because the phase
difference in the displacements of the two ions in the same unit cell.

As above, we simplify the expressions by adopting $C_{0}^{(x)}=C_{0}%
^{(y)}=C_{1}^{(x)}=C_{1}^{(y)}=C$, $C_{1}^{(z)}=2\, C$ and $C_{0}^{(z)}/C\gg
1$. the phonon dispersion now read
\begin{align}
\omega_{kO}^{(x)}  &  =\omega_{kO}^{(y)}=\sqrt{2}\omega_{0}\left[
1+\sqrt{\frac{1+\cos kb}{2}}\right]  ^{-1/2},\nonumber\\
\omega_{kO}^{(z)}  &  \approx\sqrt{\frac{C_{0}^{z}}{C}}\omega_{0}\left[
1+\sqrt{1-\frac{2C}{C_{0}^{z}}\left(  1-\cos kb\right)  }\right]  ^{-1/2}%
\gg\omega_{0},\nonumber\\
\omega_{kA}^{(x)}  &  =\omega_{kA}^{(y)}=\sqrt{2}\omega_{0}\left[
1-\sqrt{\frac{1+\cos kb}{2}}\right]  ^{-1/2},\nonumber\\
\omega_{kA}^{(z)}  &  \approx\sqrt{\frac{C_{0}^{z}}{C}}\omega_{0}\left[
1-\sqrt{1-\frac{2C_{1}^{(z)}}{C_{0}^{z}}\left(  1-\cos kb\right)  }\right]
^{-1/2}\rightarrow\sqrt{2}\omega_{0}\sqrt{1-\cos kb}. \tag{$\rm S.B6$}%
\label{padip}%
\end{align}
This approximation restores translational symmetry for the transverse modes
and the gaps at $k=\pi/b$ vanish (see Fig. \ref{Fig-SM1}). The polarization
carried by the transverse acoustic and optical phonons reads
\begin{align}
\xi_{kO}^{(\alpha)}  &  =-\frac{\hbar}{2P_{0}}\frac{|\mathbf{F}_{kO}%
^{(\alpha)}|^{2}}{\omega_{kO}^{(\alpha)}}=\frac{Q^{2}\hbar\omega_{kO}%
^{(\alpha)}}{4P_{0}C},\nonumber\\
\xi_{kA}^{(\alpha)}  &  =-\frac{\hbar}{2P_{0}}\frac{|\mathbf{F}_{kA}%
^{(\alpha)}|^{2}}{\omega_{kA}^{(\alpha)}}=\frac{Q^{2}\hbar\omega_{kA}%
^{(\alpha)}}{4P_{0}C} \tag{$\rm S.B7$}\label{papo}%
\end{align}
where $\alpha=x,y$. In contrast to the perpendicular case, the polarization
carried by the transverse optical and acoustic modes is proportional to their
frequencies. With Eq.~(\ref{padip}) and Eq.~(\ref{papo}), the transport
coefficients are then given by
\begin{align}
G  &  =-\frac{1}{\hbar}\sum_{\alpha=x,y}\left[  \int_{0}^{\sqrt{2}\omega_{0}%
}[\xi_{kA}^{(\alpha)}]^{2}\frac{\partial N(\omega_{kA}^{(\alpha)},T)}%
{\partial\omega_{kA}^{(\alpha)}}\frac{d\omega_{kA}^{(\alpha)}}{2\pi}%
+\int_{\sqrt{2}\omega_{0}}^{2\omega_{0}}[\xi_{kO}^{(\alpha)}]^{2}%
\frac{\partial N(\omega_{kO}^{(\alpha)},T)}{\partial\omega_{kA}^{(\alpha)}%
}\frac{d\omega_{kO}^{(\alpha)}}{2\pi}\right] \nonumber\\
&  =\frac{\xi_{P}^{2}}{8h}\int_{0}^{2\epsilon_{0}}d\epsilon\frac{\epsilon
^{2}e^{\epsilon}}{\epsilon_{0}^{2}(e^{\epsilon}-1)^{2}}\nonumber\\
\Pi &  =ST=G^{-1}\sum_{\alpha=x,y}\left[  \int_{0}^{\sqrt{2}\omega_{0}}%
\omega_{kA}^{(\alpha)}\xi_{kA}^{(\alpha)}\frac{\partial N(\omega_{kA}%
^{(\alpha)},T)}{\partial\omega_{kA}^{(\alpha)}}\frac{d\omega_{kA}^{(\alpha)}%
}{2\pi}+\int_{\sqrt{2}\omega_{0}}^{2\omega_{0}}\omega_{kO}^{(\alpha)}\xi
_{kO}^{(\alpha)}\frac{\partial N(\omega_{kO}^{(\alpha)},T)}{\partial
\omega_{kO}^{(\alpha)}}\frac{d\omega_{kO}^{(\alpha)}}{2\pi}\right] \nonumber\\
&  =\frac{4C\delta^{2}}{P_{0}}\nonumber\\
K  &  \approx\sum_{\alpha=x,y}\left[  \int_{0}^{\sqrt{2}\omega_{0}}\hbar
\omega_{kA}^{(\alpha)}\frac{\partial N(\omega_{kA}^{(\alpha)},T)}{\partial
T}\frac{d\omega_{kA}^{(\alpha)}}{2\pi}+\int_{\sqrt{2}\omega_{0}}^{2\omega_{0}%
}\hbar\omega_{kO}^{(\alpha)}\frac{\partial N(\omega_{kO}^{(\alpha)}%
,T)}{\partial T}\frac{d\omega_{kO}^{(\alpha)}}{2\pi}\right]  +\int%
_{0}^{2\omega_{0}}\hbar\omega_{kA}^{(z)}\frac{\partial N(\omega_{kA}^{(z)}%
,T)}{\partial T}\frac{d\omega_{kA}^{(z)}}{2\pi}\nonumber\\
&  =\frac{3k_{B}^{2}T}{h}\int_{0}^{2\epsilon_{0}}d\epsilon\frac{\epsilon
^{2}e^{\epsilon}}{(e^{\epsilon}-1)^{2}} \tag{$\rm S.B8$}%
\end{align}
where $(0,\sqrt{2}\omega_{0})$, $(0,2\omega_{0})$ and $(\sqrt{2}\omega
_{0},2\omega_{0})$ are the band edges for $Ax$ ($Ay$), $Az$ and $Ox$ ($Oy$)
modes, respectively. In contrast to the case of perpendicular configuration,
the massless transverse acoustic phonons contribute significantly to the
polarization transport. We disregarded again the contribution of the
high-frequency longitudinal optical mode $Oz$ to the heat conductance.
